\documentclass[preprint,showpacs,preprintnumbers,amsmath,amssymb]{revtex4}

\usepackage{graphicx}
\usepackage{dcolumn}
\usepackage{bm}

\begin{document}

\preprint{AS-ITP-2005-003}

\title{Supersymmetry for Fermion Masses} 

\author{Chun Liu}  
\affiliation{Institute of Theoretical Physics, 
Chinese Academy of Sciences,\\
P. O. Box 2735, Beijing 100080, China}
 \email{liuc@itp.ac.cn}

\date{\today}

\begin{abstract}
  It is proposed that supersymmetry (SUSY) maybe used to understand 
fermion mass hierarchies.  A family symmetry $Z_{3L}$ is introduced, 
which is the cyclic symmetry among the three generation $SU(2)$ 
doublets.  SUSY breaks at a high energy scale $\sim 10^{11}$ GeV.  
The electroweak energy scale $\sim 100$ GeV is unnaturally small.  No 
additional global symmetry, like the R-parity, is imposed.  The Yukawa 
couplings and R-parity violating couplings all take their natural values 
which are ${\mathcal O}(10^0-10^{-2})$.  
Under the family symmetry, only the third 
generation charged fermions get their masses.  This family symmetry is 
broken in the soft SUSY breaking terms which result in a hierarchical 
pattern of the fermion masses.  It turns out that for the charged 
leptons, the $\tau$ mass is from the Higgs vacuum expectation value 
(VEV) and the sneutrino VEVs, 
the muon mass is due to the sneutrino VEVs, and the electron 
gains its mass due to both $Z_{3L}$ and SUSY breaking.  
The large neutrino mixing are produced 
with neutralinos playing the partial role of right-handed neutrinos.  
$|V_{e3}|$ which is for $\nu_e-\nu_{\tau}$ mixing is expected to be 
about $0.1$.  For the quarks, the third generation masses are from the 
Higgs VEVs, the second generation masses are from quantum corrections, 
and the down quark mass due to the sneutrino VEVs.  It explains 
$m_c/m_s$, $m_s/m_e$, $m_d > m_u$ and so on.  Other aspects of the model 
are discussed.  
\end{abstract}



\maketitle

\section{Introduction}

In elementary particle physics, SUSY \cite{susy} was proposed for 
stabilizing the electroweak (EW) energy scale \cite{susy1,susy2} which 
is otherwise unnaturally small compared to the grand unification scale 
$3\times 10^{16}$ GeV \cite{gut,gut1}.  The study of the cosmological constant 
\cite{cc}, however, suggests that unnaturalness of $10^{120}$ 
fine tuning might be just so from the anthropic point of view.  It was 
argued that the string theory even supports the emergence of the anthropic 
landscape \cite{string}.  This led to a consideration of giving up 
naturalness of the EW scale \cite{sm,split}.  To keep gauge coupling constant 
unification and the dark matter, the so-called split SUSY \cite{split,split1} 
was invented which has new features phenomenologically 
\cite{split2,r-parity2,calmet,axion,manuel}.  

In Ref. \cite{liu4}, it was asked that if SUSY is not for stabilizing the EW 
scale, what else job does this beautifully mathematical physics do in particle 
physics, other than gauge coupling unification and the dark matter?  
We proposed to make use of SUSY to understand the lepton 
mass hierarchies.  The flavor puzzle, namely the fermion masses, mixing and CP 
violation, in the Standard Model (SM) needs new physics to be understood 
\cite{he}.  The empirical fermion mass pattern is that the third generation is 
much heavier than the second generation which is also much heavier than the 
first.  This may imply a family symmetry 
\cite{family-symmetry,liu1,liu2}.  We first considered the charged 
leptons.  By assuming a $Z_3$ cyclic symmetry among the $SU(2)$ doublets 
$L_i$ ($i=1,2,3$) of the three generations \cite{liu1,liu2}, 
only the tau lepton gets mass, the muon 
and electron are still massless.  The essential point is how the family 
symmetry breaks.  Naively the symmetry breaking can be achieved by introducing 
family-dependent Higgs fields.  We observed that SUSY naturally provides such 
Higgs-like fields, which are the scalar neutrinos.  If the vacuum expectation 
values (VEVs) of the sneutrinos are non-vanishing, $v_i\neq 0$, the R-parity 
violating interactions $L_iL_jE_k^c$ \cite{r-parity,r-parity1}, with 
$E_k^c$ denoting the anti-particle superfields of the SU(2) singlet 
leptons, contribute to the fermion masses, in addition to the Yukawa 
interactions.  This is the origin of family symmetry breaking.  The 
above idea has been proposed for some time \cite{liu1,liu2}.  Because 
SUSY had been used to stabilize the EW scale, that idea suffered from 
severe constraints.  For example, the $\tau$-neutrino should be $10$ MeV 
heavy \cite{liu3}.  It is a liberation if SUSY has nothing to do with 
the EW scale.  Because the SUSY breaking scale is very high, the 
neutrinos are light.  Furthermore, there is no need to introduce the 
R-parity or the baryon number as a symmetry.  

In this paper, after refining the lepton sector, we include discussion of the 
quark masses.  To understand the large ratio of the top quark mass and the 
bottom quark mass, we assume that $\tan \beta$ is large.  Numerically we make 
modification correspondingly.  While the $\tau$-lepton mass is from the 
down-type Higgs VEV $\sim 10$ GeV, the muon mass is due to $v_i$, 
$m_\mu\sim\lambda v_i$ with $\lambda$ standing for the trilinear 
R-parity violation couplings.  It is natural $\lambda\sim 10^{-1}$ 
like the Yukawa couplings for the $\tau$ mass.  The muon mass tells us 
then $v_i\sim 1$ GeV.  $1$ GeV $v_i$'s could induce a large lepton 
number violating effect, namely a large neutrino Majorana mass if the 
neutralinos are not heavy, due to 
$\displaystyle m_\nu\simeq (g_2 v_i)^2/M_{\tilde{Z}}$, where $g_2$ is 
the $SU(2)_L$ gauge coupling constant, and $M_{\tilde{Z}}$ is the 
gaugino mass.  When we take $M_{\tilde{Z}} \simeq 10^{11}$ GeV, the above 
formula can produce a neutrino mass needed to explain the solar 
neutrino problem.  

This paper is organized as follows.  In Sect. II, we will review, improve 
and expand the discussion of the lepton sector \cite{liu4}.  
Quark sector is 
studied in Sect. III.  In addition to the quark masses, mixing and CP 
violation are considered.  Sect. IV gives the low energy effective 
theory.  It will be easy and clear to discuss the neutrino masses and the 
lepton mixing in a separate section which is Sect. V.  Sect. VI discusses 
some important and interesting aspects of the 
model.  A summary is given in the final section.  

\section{Leptons}

In our model the $Z_{3L}$ family symmetry, that is invariance under 
$L_1 \to L_2 \to L_3 \to L_1$, mentioned in the beginning is 
assumed, which however is softly broken.  The gauge symmetries and the 
matter contents in the full theory are the same as those in the SUSY SM.  
When the family symmetry is considered, the relevant kinetic terms 
should be written in a general form which keeps the symmetry,  
\begin{equation}
\label{1}
\begin{array}{lll}
{\mathcal  L} &\supset& 
\left( H_1^{\dag}H_1+H_2^{\dag}H_2+\alpha L_i^{\dag}L_i
+\beta (L_1^{\dag}L_2+L_2^{\dag}L_3+L_3^{\dag}L_1+h.c.)\right.\\[3mm]
 & &\displaystyle 
+\left.\left. \frac{\gamma}{\sqrt{3}}(H_2^{\dag}\sum_i L_i+h.c.)\right)
\right|_{\theta\theta\bar{\theta}\bar{\theta}} \,,
\end{array}
\end{equation}
where $H_1$ and $H_2$ are the two Higgs doublets, $\alpha$, $\beta$, 
$\gamma$ are $O(1)$ coefficients.  The case of that $\alpha=1$ and 
$\beta=\gamma=0$ is a special one of above expression.  The 
superpotential is 
\begin{equation}
\label{2}
{\mathcal W} = \frac{\tilde{y}_j}{\sqrt{3}}(\sum_i L_i)H_2E^c_j 
+\tilde{\lambda}_j(L_1L_2+L_2L_3+L_3L_1)E^c_j
+\tilde{\mu}H_1H_2+\tilde{\mu}'H_1\sum_i L_i\,,
\end{equation}
where $\tilde{y}_j$'s and $\tilde{\lambda}_j$'s are the coupling 
constants.  $\tilde{\mu}$ and $\tilde{\mu}'$ are mass terms.  It is 
natural that their order closes to the scale of soft SUSY breaking 
masses.  The Lagrangian of soft SUSY breaking masses is 
\begin{equation}
\label{3}
\begin{array}{lll}
{\mathcal L}_{soft1}&=& M_{\tilde{W}}\tilde{W}\tilde{W}
             +M_{\tilde{Z}}\tilde{Z}\tilde{Z} \\[3mm] 
&&+m_h^2h_1^{\dag}h_1+m_h^2h_2^{\dag}h_2
+m_{lL_{ij}}^2\tilde{l}_i^{\dag}\tilde{l}_j
+m_{lR_{ij}}^2\tilde{e}_i^*\tilde{e}_j \\[3mm] 
&&+(B_{\tilde{\mu}} h_1h_2+B_{\tilde{\mu}_i} h_1\tilde{l_i}
+m_i^{\prime 2} h_2^{\dag}\tilde{l}_i+h.c.)\,,
\end{array}
\end{equation}
where $\tilde{W}$ and $\tilde{Z}$ stand for the charged and neutral 
gauginos, respectively,  $h_1$, $h_2$, $\tilde{l}_i$ and $\tilde{e}_i$ 
are the scalar components of $H_1$, $H_2$, $L_i$ and $E^c_i$ 
respectively.  Note that explicitly breaking of $Z_{3L}$ is introduced 
in the soft mass terms.   The soft masses are assumed to be very large 
around a typical mass $m_S$.  The trilinear soft terms should be also 
included, 
\begin{equation}
\label{4}
{\mathcal L}_{soft2}=\tilde{m}_{ij}\tilde{l}_ih_2\tilde{e}_j
+\tilde{m}_{ijk}\tilde{l}_i\tilde{l}_j\tilde{e}_k+h.c.\,.  
\end{equation}
The mass coefficients which we denote generally as $\tilde{m}_S$ can 
be close to $m_S$.  

The expression of the kinetic terms is not yet in the normalized 
canonical form.  The standard form 
\begin{equation}
\label{5}
{\mathcal L} \supset \left. \left(
H_u^{\dag} H_u+H_d^{\prime\dag}H_d'+L_e^{\dag}L_e
+L_{\mu}^{\dag}L_{\mu}+L_{\tau}^{\prime\dag}L_{\tau}' \right)
\right|_{\theta\theta\bar{\theta}\bar{\theta}}   
\end{equation}
is achieved by the field re-definition: 
\begin{equation}
\label{6}
\begin{array}{lll}
H_u    &=& H_1 \,, \\
H_d'   &=& \displaystyle 
c_1 \left(H_2+\frac{c_2}{\sqrt{3}}\sum_i L_i\right) \,, \\[3mm]
L_{\tau}'&=& \displaystyle 
c_1'\left(H_2-\frac{c_2}{\sqrt{3}}\sum_i L_i\right) \,, \\[3mm]
L_{\mu}  &=& \displaystyle \frac{c_3}{\sqrt{2}}(L_1-L_2)\cos\theta
          +\frac{c_3}{\sqrt{6}}(L_1+L_2-2L_3)\sin\theta \,, \\[3mm]
L_e    &=&-\displaystyle \frac{c_3}{\sqrt{2}}(L_1-L_2)\sin\theta
          +\frac{c_3}{\sqrt{6}}(L_1+L_2-2L_3)\cos\theta \,,
\end{array}
\end{equation}
where 
\begin{equation}
\label{7}
c_1 =\displaystyle\frac{1}{\sqrt{2}}\sqrt{1+\frac{\gamma}{c_2}} \,, ~~
     c_2=\sqrt{\alpha+2\beta} \,, ~~ c_3=\sqrt{\alpha-\beta} \,, ~~
c_1'=\displaystyle \frac{1}{\sqrt{2}}\sqrt{1-\frac{\gamma}{c_2}} \,
\end{equation}
and $\theta$ can not be determined until muon mass basis is fixed.  

The superpotential is then 
\begin{equation}
\label{8}
{\mathcal W} = \sqrt{\sum_j |y_j|^2} H_d'L_{\tau}' E^c_{\tau}  
+L_eL_{\mu}(\lambda_{\tau} E^c_{\tau}+\lambda_{\mu} E^c_{\mu}) 
+\mu H_uH_d'+ \mu' H_u L_{\tau}' \,,
\end{equation}
where 
\begin{equation}
\label{9}
\begin{array}{llll}
y_j & = & \displaystyle\frac{2}{\sqrt{\alpha+2\beta-\gamma^2}}
          \tilde{y_j}\,, ~~& 
\lambda_j = \displaystyle -\frac{\sqrt{3}}{\alpha+\beta}
\tilde{\lambda_j}\,,\\[3mm]
\mu & = & \displaystyle \frac{1}{2c_1}\left(\tilde{\mu}
+\frac{\tilde{\mu}'}{c_2}\right)\,, ~~ &
\mu' = \displaystyle \frac{1}{2c_1'}\left(\tilde{\mu}
-\frac{\tilde{\mu}'}{c_2}\right)\,,
\end{array}
\end{equation}
$E^c_{\tau}$ is defined as 
\begin{equation}
\label{10}
E^c_{\tau} = \frac{y_j}{y_{\tau}} E^c_j\,, 
\end{equation} 
where $y_{\tau}\equiv \sqrt{\sum_j |y_j|^2}$.  
$E^c_{\mu}$ is orthogonal to $E^c_{\tau}$, $\lambda_{\tau}$ and 
$\lambda_{\mu}$ are combinations of $y_j$'s and $\lambda_j$'s.  Because 
of the $Z_{3L}$ symmetry, the superpotential is without the field 
$E^c_e$ which is orthogonal to both $E^c_{\tau}$ and $E^c_{\mu}$.  

To look at the fermion masses, we simply rotate the bilinear R-parity 
violating term away via the field re-definition, 
\begin{equation}
\label{11}
H_d = \displaystyle\frac{1}{\bar{\mu}}(\mu H_d'+\mu'L_{\tau}')\,,~~ 
L_{\tau} = \displaystyle 
           \frac{1}{\bar{\mu}}(\mu' H_d'-\mu L_{\tau'})\,,  
\end{equation}
where $\bar{\mu}\equiv\sqrt{\mu^2+\mu^{\prime 2}}$.  
It is trivial to see that the kinetic terms are diagonal in terms of 
$H_d$ and $L_{\tau}$.  The superpotential is 
\begin{equation}
\label{12}
{\mathcal W} = -y_{\tau} H_d L_{\tau} E^c_{\tau}  
+L_eL_{\mu}(\lambda_{\tau} E^c_{\tau}+\lambda_{\mu} E^c_{\mu})  
    +\bar{\mu} H_uH_d \,.  
\end{equation}
The $Z_{3L}$ family symmetry keeps the trilinear R-parity violating 
terms invariant.  As we have expected Higgs field $H_d$ contributes to 
the tau mass only and the sneutrinos in $L_e$ and $L_{\mu}$ contribute 
to the muon mass, after they get VEVs.  The VEVs of $L_e$ and $L_{\mu}$ 
imply the breaking of the $Z_{3L}$ symmetry as can be seen explicitly 
from Eq. (\ref{6}).  The electron remains massless because of absence of 
the $E^c_e$ field in ${\mathcal W}$.  A hierarchy among charged leptons 
is obtained.  

The breaking of the family symmetry originates from the soft SUSY 
masses.  For simplicity and without losing generality, we assume 
that the soft terms in Eqs. (\ref{3}) and (\ref{4}) are rewritten as 
\begin{equation}
\label{13}
\begin{array}{lll}
{\mathcal L}_{soft}&=& M_{\tilde{W}}\tilde{W}\tilde{W}
             +M_{\tilde{Z}}\tilde{Z}\tilde{Z} \\[3mm] 
&&+m_{h_u}^2h_u^{\dag}h_u+m_{h_d}^2h_d^{\dag}h_d
+m_{h_d}^2\tilde{l}_{\alpha}^{\dag}\tilde{l}_{\alpha}
+m_{lR_{\alpha\beta}}^2\tilde{e}^*_{\alpha} \tilde{e}_{\beta} \\[3mm] 
&&+(B_{\mu} h_uh_d+B_{\mu_{\alpha}} h_u\tilde{l}_{\alpha} 
+\tilde{m}_{\alpha\beta}\tilde{l}_{\alpha} h_d\tilde{e}_{\beta}
+\tilde{m}_{\alpha\beta\gamma}
\tilde{l}_{\alpha}\tilde{l}_{\beta}\tilde{e}_{\gamma}+h.c.)\,, 
\end{array}
\end{equation}
where $\alpha=e, \mu, \tau$.  

The key point of the form of the scalar masses lies in the 
$(h_u~~h_d^{\dag}~~\tilde{l}_{\alpha}^{\dag})$ mass-squared matrix, 
\begin{equation}
\label{14}
{\mathcal M}^{(h_u,h_d^{\dag},\tilde{l}_{\alpha}^{\dag})} = 
\left(
\begin{array}{ccccc}
m_{h_u}^2         &B_{\mu}   &B_{\mu_e} &B_{\mu_{\mu}}&B_{\mu_{\tau}}\\
B_{\mu}       &m_{h_d}^2 &0         &0            &0             \\
B_{\mu_e}     &0         &m_{h_d}^2 &0            &0             \\
B_{\mu_{\mu}} &0         &0         &m_{h_d}^2    &0             \\
B_{\mu_{\tau}}&0         &0         &0            &m_{h_d}^2     \\
\end{array}
\right) 
\end{equation}
of which the eigenvalues are 
\begin{equation}
\label{15}
\begin{array}{lll}
M_1^2 & = & \displaystyle\frac{m_{h_u}^2+m_{h_d}^2}{2}-
\sqrt{\displaystyle\left(\frac{m_{h_u}^2-m_{h_d}^2}{2}\right)^2
+(B_{\mu})^2+\sum_{\alpha}(B_{\mu_{\alpha}})^2} \\
M_2^2 & = & \displaystyle\frac{m_{h_u}^2+m_{h_d}^2}{2}+
\sqrt{\displaystyle\left(\frac{m_{h_u}^2-m_{h_d}^2}{2}\right)^2
+(B_{\mu})^2+\sum_{\alpha}(B_{\mu_{\alpha}})^2} \\
M_3^2 & = & M_4^2 = M_5^2 = m_{h_d}^2\,.  
\end{array}
\end{equation}
It is understood that $m_{h_u}^2$ and $m_{h_d}^2$ appeared in the 
matrix Eq. (\ref{14}) denote the sum of the squared soft masses and the 
squared masses generated from the superpotential.  $m_{h_u}^2$ can be 
negative.  The analysis goes in the similar way as in Ref. \cite{split}.  
By fine-tuning, $M_1^2\sim -m_{EW}^2$, namely the EW symmetry breaking 
is achieved.  The tuning is at the order of $m_S^2/m_{EW}^2$.  

In our case, in addition to the Higgs doublets, $\tilde{l}_{\alpha}$ 
fields also get VEVs, 
\begin{equation}
\label{16}
v_u \neq 0\,, ~~v_d\neq 0\,, ~~ v_{l_{\alpha}} \neq 0 ~~~
(\alpha = e,~\mu,~ \tau)\,.  
\end{equation}
The relative size of these values are determined by the soft mass 
parameters.  It is easy to show from Eqs. (\ref{14}) and (\ref{15}) 
that $v_{l_{\alpha}}/v_d = B_{\mu_{\alpha}}/B_{\mu}$ and 
$v_{l_{\alpha}}/v_{l_{\beta}} = B_{\mu_{\alpha}}/B_{\mu_{\beta}}$.  
It is therefore possible that hierarchies among 
$v_{u}$, $v_{d}$ and $v_{l_{\alpha}}$ occur if there are hierarchies 
among the $B_{\mu}$'s.  
Note that the $L_{\alpha}$ numbers break explicitly in the soft mass 
terms, nonvanishing $v_{l_{\alpha}}$'s do not result in any massless 
scalar.  Because there is only one light Higgs doublet, the tree-level 
flavor changing neutral current (FCNC) does not appear.  The 
hierarchical charged lepton mass pattern is obtained from 
Eq. (\ref{12}) explicitly, 
\begin{equation}
\label{17}
m_{\tau} \sim y_{\tau} v_d \,,~ 
m_{\mu}  \sim \lambda_{\mu} \sqrt{v_{l_e}^2+v_{l_{\mu}}^2} \,, ~ 
m_e      =    0 \,. 
\end{equation}
Numerically it is required that $v_d\sim 10$ GeV and 
$\sqrt{v_{l_e}^2+v_{l_{\mu}}^2} \sim 1$ GeV.  
A careful analysis will be given in Sect. V.  

\begin{figure}
\includegraphics{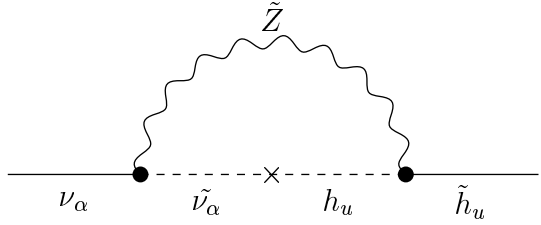}
\caption{\label{fig1}
$B_{\mu_{\alpha}}$ induces a large lepton-Higgsino mixing via one loop.}
\end{figure}

It is important to note that masslessness of the electron is kept by SUSY.  
Generally, family symmetries keep the muon and electron massless.  Once the 
family symmetry is broken, however, both muon and electron get their masses.  
And there is no reason to expect a hierarchy between the muon mass and the 
electron mass.  In this model, it is the simplicity of the superpotential 
Eq. (\ref{8}) that makes the electron massless even if the sneutrino VEVs are 
non-vanishing.  The simplicity comes from SUSY.  The non-vanishing electron 
mass is therefore due to SUSY breaking effects, as will be seen later.  

If large $v_{l_{\alpha}}$'s are safe should be studied.  In addition, 
it should be also considered that huge $B_{\mu_{\alpha}}$'s induce large 
lepton-Higgsino mixing.  The inducement happens at the loop-level 
through the gaugino exchange, as shown in Fig.~\ref{fig1} \cite{liu3}, 
$m_{\alpha h}=\displaystyle\frac{g_2^2B_{\mu_{\alpha}}}{16\pi^2M_{\tilde{Z}}}$ 
which is about $10^{-3}m_S$.  By denoting $\tilde{h}$ as Higgsinos, 
the mass matrix of $\nu_{\alpha}$ and the other neutralinos is given as 
\begin{equation}
\label{18}
-i\left(\nu_e~\nu_{\mu}~\nu_{\tau}~\tilde{h}_d^0~\tilde{h}_u^0~\tilde{Z}
\right)
\left(
\begin{array}{cccccc}
0       & 0          &0            &0         & m_{eh}   &av_{l_e}    \\
0       & 0          &0            &0         &m_{\mu h} &av_{l_{\mu}}\\
0       & 0          &0            &0         &m_{\tau h}&av_{l_{\tau}}\\
0       & 0          &0            &0         &-\bar{\mu}&av_d \\
m_{eh}  &m_{\mu h}   &m_{\tau h}   &-\bar{\mu}&0         &-av_u\\
av_{l_e}&av_{l_{\mu}}&av_{l_{\tau}}& av_d     &-av_u     &M_{\tilde{Z}}
\end{array}
\right)
\left(
\begin{array}{c}
\nu_e\\ \nu_{\mu}\\ \nu_{\tau}\\ \tilde{h}_d^0\\ \tilde{h}_u^0\\ \tilde{Z} 
\end{array}
\right)\,,
\end{equation}
where $a=\displaystyle(\frac{g_2^2+g_1^2}{2})^{1/2}$ with $g_1$ being 
the SM $U(1)_Y$ coupling constant.  We simply obtain the three large mass 
eigenvalues of the above mass matrix by reasonably taking 
$v_{l_{\alpha}},~ v_d,~ v_u\ll \bar{\mu}, M_{\tilde{Z}}$, 
\begin{equation}
\label{19}
\Lambda_1 \simeq M_{\tilde{Z}} \,,~~ \Lambda_2 \simeq \bar{\mu} \,,~~ 
\Lambda_3 \simeq -\bar{\mu} \,.
\end{equation}  
For the three light neutrinos, an interesting observation is that the 
mass matrix Eq. (\ref{18}) is a realization of the see-saw mechanism 
\cite{seesaw}.  In the above mass matrix, we denote $M_R$ being the 
$3 \times 3$ lower-right submatrix, and $m_{\rm Dirac}$ the $3 \times 3$ 
upper-right submatrix.  The heavy higgsinos and gauginos play the role of 
the right-handed neutrinos, a $3 \times 3$ light Majorana neutrino mass 
matrix is obtained as 
\begin{equation}
\label{20}
\begin{array}{lll}
{\mathcal M}_0^{\nu}&\simeq&-m_{\rm Dirac} M_R^{-1} m_{\rm Dirac}^T \,, \\
[3mm]
& = & \displaystyle - \frac{a^2}{M_{\tilde{Z}}}
\left(
\begin{array}{ccc}
v_{l_e}v_{l_e}      &v_{l_e}v_{l_{\mu}}      &v_{l_e}v_{l_{\tau}}     \\
v_{l_{\mu}}v_{l_e}  &v_{l_{\mu}}v_{l_{\mu}}  &v_{l_{\mu}}v_{l_{\tau}} \\
v_{l_{\tau}}v_{l_e} &v_{l_{\tau}}v_{l_{\mu}} &v_{l_{\tau}}v_{l_{\tau}}
\end{array}
\right) \,.
\end{array}  
\end{equation}
The mass matrix is of rank $1$.  The nonvanishing mass is 
$m_{\nu}=\displaystyle\frac{a^2}{M_{\tilde{Z}}}
v_{l_{\alpha}}v_{l_{\alpha}}$.  It is very small $\sim 10^{-1}-10^{-3}$ 
eV when $M_{\tilde{Z}}\sim 10^{9}-10^{11}$ GeV and 
$v_{l_{\alpha}}\sim (1-10)$ GeV.   By 
introducing right-handed neutrinos, the neutrino sector has freedom 
to accommodate the realistic neutrino oscillation data.  

We note that in the superpotential Eq. (\ref{12}), the lepton number is 
violated.  However, this violation is suppressed by gaugino and slepton 
masses, it has no observable effects at low energies.  For example, the 
loop-induced electron-neutrino mass due to the R-parity violating 
trilinear interactions \cite{liu1} is 
$m_{\nu_e}\simeq\displaystyle\frac{\lambda_{\mu}^2}{16\pi^2}
\frac{\lambda_{\mu}v_dm_{\mu}}{m_S}\sim 10^{-5}-10^{-6}$ eV.  This is 
too small to be relevant to current neutrino physics.  

\begin{figure}
\includegraphics{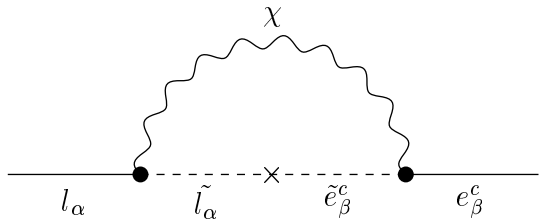}
\caption{\label{fig2}
SUSY loop generation of the charged lepton masses.  $\chi$ and $l$,
$e^c$ denote the neutral gauginos and charged leptons.}
\end{figure}

The electron mass comes from the loop effects of $Z_{3L}$ violation in 
the soft terms \cite{liu1}.  The soft breaking of $Z_{3L}$ generates 
non-vanishing masses for the charged leptons through the one loop 
diagram Fig.~\ref{fig2}, where $\chi$ and $l$, $e^c$ denote the neutral 
gauginos and charged leptons.  The mixing of the scalar leptons 
associated with different chiralities is due to the soft trilinear 
terms in Eq. (\ref{13}), which is then about $y_{\tau}\tilde{m}_S v_d$.  
The exact formula for the one loop induced masses is 
\begin{equation}
\label{22a}
\delta M^l_{\alpha\beta}=\sum_{\chi}\frac{g_{\chi}^2}{16\pi^2}
\frac{m_{\chi}}{m_{\chi}^2-m_{\tilde{l}_{\beta}^c}^2}
\left(\frac{m_{\chi}^2}{m_{\chi}^2-m_{\tilde{l}_{\alpha}}^2}
\ln\frac{m_{\tilde{l}_{\alpha}}^2}{m_{\chi}^2}+
\frac{m_{\tilde{l}_{\beta}^c}^2}{m_{\tilde{l}_{\alpha}}^2
-m_{\tilde{l}_{\beta}^c}^2}\ln\frac{m_{\tilde{l}_{\alpha}}^2}
{m_{\tilde{l}_{\beta}^c}^2}\right) y_{\tau}\tilde{m}_S v_d\,.
\end{equation}
Approximately it is 
\begin{equation}
\label{23}
\delta M^l_{\alpha\beta}\simeq \frac{\alpha}{\pi}
\frac{y_{\tau}\tilde{m}_S v_d}{m_S} \,.  
\end{equation}
Taking $\tilde{m}_S/m_S\simeq 0.1$, 
$\delta M^l_{\alpha\beta} \sim {\cal O}$(MeV) which determines the 
electron mass.  Note that the loop induced SUSY breaking effects are 
suppressed by the high SUSY breaking scale $m_S$ in our case.  This is 
different from the split SUSY case where 
$m_{\chi}/m_{\tilde{l}_{\alpha}^{(c)}}\to 0$.  

The loop in Fig. 2 does not cause any SUSY FCNC problem.  When a photon 
line is attached to the internal lines, the amplitude of the FCNC 
process is suppressed by a factor $v_d/m_S$ which in our scenario is 
unobservably small $\sim 10^{-9}-10^{-10}$.  Therefore the loop 
generates masses only.  In effective theory language, it produces a 
purely ordinary SM Yukawa interaction.  This point is different from the 
weak scale SUSY \cite{ah}.  Actually, whenever the SUSY breaking scale 
is pushed arbitrarily high, while keeping a Higgs unnaturally light, the 
radiative fermion mass generation mechanism is viable.  

\section{Quarks}

Now let us come to the quark masses.  Like that of the charged leptons, 
the quark masses also have three origins: the Higgs VEVs, the sneutrino 
VEV and the loop effects of the soft $Z_{3L}$ violating terms.  However, 
the roles of the sneutrino VEVs and the loop effects are switched 
\cite{liu2}.  The sneutrino VEVs contribute to the first generation 
quark masses, and the loop effects to the charm and strange quark masses.  
Under the family symmetry $Z_{3L}$, the three quark $SU(2)$ doublets 
$Q_i$ are also cyclic.  The $Z_{3L}$ symmetric superpotential includes 
\begin{equation}
\begin{array}{lll}
\label{3.1}
{\mathcal W} &\supset& \displaystyle\frac{y^u_j}{\sqrt{3}}
(\sum_i Q_i)H_1U^c_j+ \frac{y^d_j}{\sqrt{3}}(\sum_i Q_i)H_2 D^c_j
   +\lambda'_{1j}\sum_iQ_iL_iD^c_j \\[3mm] 
& &+\lambda'_{2j}(Q_1L_2+Q_2L_3+Q_3L_1)D^c_j
   +\lambda'_{3j}(Q_1L_3+Q_2L_1+Q_3L_2)D^c_j\,, 
\end{array}
\end{equation}
where $U^c_i$ and $D^c_i$ are the $SU(2)$ singlet superfields for the 
up- and down-type quarks, respectively.  $y^{u(d)}_j$ and 
$\lambda'_{ij}$ are coupling constants.  The new soft terms are masses 
of the squarks and the trilinear terms corresponding to Eq. (\ref{3.1}), 
but without $Z_{3L}$ symmetry.  The kinetic terms include 
\begin{equation}
\label{3.2} 
{\mathcal  L} \supset \left. \left[\alpha' Q_i^{\dag}Q_i 
+\beta' (Q_1^{\dag}Q_2+Q_2^{\dag}Q_3+Q_3^{\dag}Q_1+h.c.)\right] 
\right|_{\theta\theta\bar{\theta}\bar{\theta}} \,, 
\end{equation}
with $\alpha'$ and $\beta'$ ${\cal O}(1)$ being coefficients.  They are 
in the canonical form 
\begin{equation}
\label{3.3} 
{\mathcal L}\supset\left.\left(Q_t^{\dag}Q_t+Q_c^{\dag}Q_c+Q_u^{\dag}Q_u
\right)\right|_{\theta\theta\bar{\theta}\bar{\theta}}\ 
\end{equation}
by the following field redefinition,
\begin{equation}
\label{3.4}
\begin{array}{lll}
Q_t&=&\displaystyle\frac{\sqrt{\alpha'+2\beta'}}{\sqrt{3}}\sum_i Q_i\\[3mm]
Q_c&=&\displaystyle \frac{c_3'}{\sqrt{2}}(Q_1-Q_2)\cos\theta'
      +\frac{c_3'}{\sqrt{6}}(Q_1+Q_2-2Q_3)\sin\theta' \\[3mm]
Q_u&=&\displaystyle -\frac{c_3'}{\sqrt{2}}(Q_1-Q_2)\sin\theta' 
      +\frac{c_3'}{\sqrt{6}}(Q_1+Q_2-2Q_3)\cos\theta' \,,
\end{array}
\end{equation}
where $c_3'=\sqrt{\alpha'-\beta'}$, $\theta'$ is still an arbitrary 
parameter.  The superpotential is then 
\begin{equation}
\label{3.5}
\begin{array}{lll}
{\mathcal W} &\supset& y_t Q_t H_u U^c_t + y_b Q_t H_d D^c_b
+Q_tL_{\tau}\sum_{\beta=b,s,d}\lambda'_{t\beta}D^c_{\beta} \\[3mm]
& & +(Q_cL_e-Q_uL_{\mu})\sum_{\beta=b,s,d}\lambda'_{c\beta}D^c_{\beta} 
+(Q_uL_e+Q_cL_{\mu})\sum_{\beta=b,s,d}\lambda'_{u\beta}D^c_{\beta} \,,
\end{array}
\end{equation}
where 
\begin{equation}
\label{3.6}
\begin{array}{lll}
y_t & = & \displaystyle\frac{1}{c_2'}\sqrt{\sum_i|y^u_i|^2}\,,~~~ 
U^c_t=\frac{y^u_i}{\sqrt{\sum_j|y^u_j|^2}}U^c_i \,,\\[3mm] 
y_b & = & \displaystyle\frac{1}{c_2'}\sqrt{\sum_i|\bar{y}^d_i|^2}\,,~~~ 
D^c_b=\frac{\bar{y}^d_i}{\sqrt{\sum_j|\bar{y}^d_j|^2}}D^c_i \,,\\[3mm] 
\bar{y}^d_i&=&\displaystyle\frac{1}{2\bar{\mu}}\left[
\left(\frac{\mu}{c_1}+\frac{\mu'}{c_1'}\right)y^d_i+\frac{1}{c_1}\left(
\frac{\mu}{c_2}-\frac{\mu'}{c_1'}\right)\sum_j\lambda'_{ji}\right]\,,
\\[5mm]
\lambda'_{\alpha b}&=&\displaystyle 
\frac{\sum_i\bar{\lambda}'_{\alpha i} \bar{y}^d_i}
{\sqrt{\sum_j|\bar{y}^d_j|^2}}~~~\mbox{for $\alpha =t,c,u$}\,,\\[5mm] 
\bar{\lambda}'_{ti}&=&\displaystyle
\frac{1}{2c_2'\bar{\mu}}\left[\left(\frac{\mu'}{c_1}
-\frac{\mu}{c_1'}\right)y^d_i+\frac{1}{c_1}\left(\frac{\mu'}{c_2}
+\frac{\mu}{c_1'}\right)\sum_j\lambda'_{ji}\right]\,,\\[3mm]
\bar{\lambda}'_{ci}&=&\displaystyle \frac{1}{2c_3c_3'}
[-(\sqrt{3}\cos\theta'-\sin\theta')(\lambda'_{1i}-\lambda'_{2i})
+(\sqrt{3}\cos\theta'+\sin\theta')(\lambda'_{1i}-\lambda'_{3i})]\,,\\[3mm]
\bar{\lambda}'_{ui}&=&\displaystyle \frac{1}{2c_3c_3'}
[(\sqrt{3}\sin\theta'+\cos\theta')(\lambda'_{1i}-\lambda'_{2i})
-(\sqrt{3}\sin\theta'-\cos\theta')(\lambda'_{1i}-\lambda'_{3i})]\,,  
\end{array}
\end{equation}  
and $\lambda'_{\alpha s}D^c_s+\lambda'_{\alpha d}D^c_d$ is that of 
$\bar{\lambda}'_{\alpha i}D^c_i$, which is orthogonal to $D^c_b$.  

Note that if we include 
$U_1^c\rightarrow U_2^c \rightarrow U_3^c \rightarrow U_1^c$ cyclic 
symmetry under the $Z_{3L}$, the above discussion does not change.  

We see that if only the Higgs fields $H_u$ and $H_d$ get VEVs, the top 
quark and bottom quark are massive which will be denoted as $m^t_0$ and 
$m^b_0$, respectively, and the other quarks are massless.  Once the 
sneutrino $\tilde{\nu}_{\alpha}$ has a VEV, additional masses contribute 
to the down-type quarks.  In the flavor basis given in Eq. (\ref{3.5}), 
it is interesting to note that the R-parity violating couplings relevant 
only to the first two generations, which can be reduced to  
$\bar{\lambda}'_{cj}$ and $\bar{\lambda}'_{uj}$, do not involve any 
Yukawa coupling $y^d_k$.  Therefore it will be natural if we take 
$\lambda'_{ij}\ll y^d_j$.  

The soft breaking terms contribute masses to the quarks via loops 
\cite{banks,liu2}.  The way is the same as that producing the electron 
mass in Fig. 2, except for that the leptons are replaced by quarks, 
and neutralinos by the gluinos $\tilde{g}$, 
\begin{equation}
\label{3.7}
\delta M^{u(d)}_{\alpha\beta}=\frac{\alpha_s}{\pi}
\frac{2m_{\tilde{g}}}{m_{\tilde{g}}^2-m_{\tilde{q}^c_{\beta}}^2}
\left(\frac{m_{\tilde{g}}^2}{m_{\tilde{g}}^2-m_{\tilde{q}_{\alpha}}^2}
\ln\frac{m_{\tilde{q}_{\alpha}}^2}{m_{\tilde{g}}^2}+
\frac{m_{\tilde{q}_{\beta}^c}^2}{m_{\tilde{q}_{\alpha}}^2
-m_{\tilde{q}_{\beta}^c}^2}\ln\frac{m_{\tilde{q}_{\alpha}}^2}
{m_{\tilde{q}_{\beta}^c}^2}\right)
(\tilde{m}_S^{u(d)})_{\alpha\beta}v_{u(d)}\,.
\end{equation}
In order to make this contribution to be for the second generation 
only, we simply assume that the trilinear soft terms are independent 
on $\alpha$ (that is they keep the $Z_{3L}$ symmetry) 
$(\tilde{m}_S)_{\alpha\beta}=\tilde{m}_{S_{\beta}}$, 
and that $m_{\tilde{q}^c}\ll m_{\tilde{q}}\sim m_{\tilde{g}}$, 
\begin{equation}
\label{3.8}
\delta M^{u(d)}_{\alpha\beta}=\frac{2\alpha_s}{\pi}
\frac{m_{\tilde{g}}^2}{m_{\tilde{g}}^2-m_{\tilde{q}_{\alpha}}^2}
\ln\frac{m_{\tilde{q}_{\alpha}}^2}{m_{\tilde{g}}^2}
\frac{\tilde{m}_{S_{\beta}}^{u(d)} v_{u(d)}}{m_{\tilde{g}}} \,.
\end{equation}
The point is that $\delta M_{\alpha\beta}$ is factorisable, 
\begin{equation}
\label{3.9}
\delta M^q_{\alpha\beta}= f^q_{\alpha}\tilde{m}^q_{\beta} \,,
\end{equation}
where $q$ stands for $u$ or $d$.  $f^q_{\alpha}$ is a function of 
$m_{\tilde{g}}$ and $m_{\tilde{q}_{\alpha}}$, 
$f^{u(d)}_{\alpha}=\displaystyle\frac{m_{\tilde{g}}^2}
{m_{\tilde{g}}^2-m_{\tilde{q}_{\alpha}}^2}
\ln\frac{m_{\tilde{q}_{\alpha}}^2}{m_{\tilde{g}}^2}$, 
$\tilde{m}^{u(d)}_{\beta}=\displaystyle
\frac{\tilde{m}_{S_{\beta}}^{u(d)} v_{u(d)}}{m_{\tilde{g}}}$.  
Neglecting $\langle\tilde{\nu}_{\alpha}\rangle$'s, 
the mass matrix is 
\begin{equation}
\label{3.10}
{\mathcal M}^{u(d)}_{\alpha\beta} = \left( \begin{array}{ccc}
f^q_{\alpha_1}\tilde{m}^q_{\beta_1}&f^q_{\alpha_1}
\tilde{m}^q_{\beta_2}&f^q_{\alpha_1}\tilde{m}^q_{\beta_3}\\
f^q_{\alpha_2}\tilde{m}^q_{\beta_1}&f^q_{\alpha_2}
\tilde{m}^q_{\beta_2}&f^q_{\alpha_2}\tilde{m}^q_{\beta_3}\\
f^q_{\alpha_3}\tilde{m}^q_{\beta_1}&f^q_{\alpha_3}
\tilde{m}^q_{\beta_2}&f^q_{\alpha_3}\tilde{m}^q_{\beta_3}+m^{t(b)}_0\\ 
\end{array} \right)\,,
\end{equation}
with $\alpha_i$ and $\beta_i$ being $(u,c,t)$ for $q$ being $u$, and $(d,s,b)$ 
for $q$ being $d$.  The mass matrix is of rank $2$.  Thus, at this stage, the 
second family quarks acquire masses.  The first family remains massless.  The 
above mass matrix determines the eigenvalues to the first order of 
$f^q\tilde{m}^q/m^{t(b)}_0$, 
\begin{equation}
\label{3.11}
\begin{array}{lll}
m_{t(b)} & \simeq & m^{t(b)}_0 \displaystyle 
\left[1+\Re\frac{f^{u(d)}_{\alpha_3}\tilde{m}^{u(d)}_{\beta_3}}
{m^{t(b)}_0}\right] \,, \\[3mm]
m_{c(s)} & \simeq & \displaystyle 
\sqrt{(|f^{u(d)}_{\alpha_1}|^2+|f^{u(d)}_{\alpha_2}|^2)
(|\tilde{m}^{u(d)}_{\beta_1}|^2+|\tilde{m}^{u(d)}_{\beta_2}|^2)}
\left[1-\frac{2\Re (f^{u(d)}_{\alpha_3}\tilde{m}^{u(d)}_{\beta_3})}
{m^{t(b)}_0}\right] \,, \\[3mm]
m_{u(d)} & =    & 0                        \,.
\end{array}
\end{equation}
The order of magnitude of the masses of the second family can be
understood naturally.  From Eq. (\ref{3.7}), we see that the charm quark 
to the strange quark mass ratio $m_c/m_s$ is mainly determined by the 
ratio $(\tilde{m}_S^uv_u)/(\tilde{m}_S^dv_d)$ if there is no significant 
difference between the masses of the squarks with same chirality.  The 
ratio $(\tilde{m}^uv_u)/(\tilde{m}^dv_d)$ can be 
$m_t/m_b\sim {\mathcal O}(10)$.  Therefore the large ratio of $m_c/m_s$ 
can be considered as a result of $m_t/m_b$.  It should also be noted that 
in Eq. (\ref{3.7}), we have neglected the other neutral gauginos, photino 
and Zino, because their effects are rather small compared with that of 
gluinos for the following two reasons.  One is that $\alpha_s$ is large, 
$\alpha_s/\alpha\sim {\mathcal O}(10)$; another is that the number of 
gluinos is $8$ which is also large.  Hence the contribution of gluinos is 
nearly two orders of magnitude larger than that of photino or Zino.  The 
radiative mass generation picture of quarks discussed above is consistent 
with that of leptons of Eq. (\ref{22a}) where it is the electron mass 
that is generated at the one-loop level by exchanging photino and Zino.  
The fact that the strange quark is two orders of magnitude heavier than 
the electron is thus explainable.  

The quark mixing are then obtained.  The mass-squared matrix 
$M^qM^{q\dag}$ is diagonalized by 
\begin{equation}
\label{3.12}
\left( \begin{array}{ccc}
  \frac{f_{\alpha_2}^{q*}}{\bar{f}^q}
&-\frac{f_{\alpha_1}^{q*}}{\bar{f}^q}&0\\[3mm]
 \frac{f_{\alpha_1}^q}{\bar{f}^q}
&\frac{f_{\alpha_2}^q}{\bar{f}^q}
&\frac{\bar{f}^q\tilde{m}_{\beta_3}^q}{m^{t(b)}_0}
\left[1+\frac{f_{\alpha_3}^q}{a}
\frac{|\tilde{m}_{\beta_1}^q|^2+|\tilde{m}_{\beta_2}^q|^2
-|\tilde{m}_{\beta_3}^q|^2}{\tilde{m}_{\beta_3}^q}\right]\\[3mm]
\frac{f_{\alpha_1}^q\tilde{m}_{\beta_i}^{q*}\tilde{m}_{\beta_i}^q}
{m^{t(b)}_0\tilde{m}_{\beta_3}^q}
\left(1-\frac{f_{\alpha_3}^q\tilde{m}_{\beta_i}^{q*}\tilde{m}_{\beta_i}^q}
{m^{t(b)}_0\tilde{m}_{\beta_3}^q}\right)
&\frac{f_{\alpha_2}^q\tilde{m}_{\beta_i}^{q*}\tilde{m}_{\beta_i}^q}
{m^{t(b)}_0\tilde{m}_{\beta_3}^q}
\left(1-\frac{f_{\alpha_3}^q\tilde{m}_{\beta_i}^{q*}\tilde{m}_{\beta_i}^q}
{m^{t(b)}_0\tilde{m}_{\beta_3}^q}\right)
&1 
\end{array} \right)\,,
\end{equation}
where $\bar{f}^q\equiv\sqrt{|f_{\alpha_1}^q|^2+|f_{\alpha_2}^q|^2}$ for 
$q$ being $u$ or $d$.  The quark mixing matrix $V_{\rm CKM}$ is 
\begin{equation}
\label{3.13}
\left( \begin{array}{ccc}
 \frac{f_{\alpha_1}^uf_{\alpha_1}^{d*}+f_{\alpha_2}^uf_{\alpha_2}^{d*}}
{\bar{f}^u\bar{f}^d}
& \frac{-f_{\alpha_1}^{u*}f_{\alpha_2}^{d*}
+f_{\alpha_2}^{u*}f_{\alpha_1}^{d*}}{\bar{f}^u\bar{f}^d}
&\frac{f_{\alpha_1}^{u*}f_{\alpha_2}^{d*}
-f_{\alpha_2}^{u*}f_{\alpha_1}^{d*}}{\bar{f}^d}
\frac{\tilde{m}_{\beta_i}^{u*}\tilde{m}_{\beta_i}^u}
{m^t_0\tilde{m}_{\beta_3}^u}\\[3mm]
\frac{f_{\alpha_1}^uf_{\alpha_2}^d-f_{\alpha_2}^uf_{\alpha_1}^d}
{\bar{f}^u\bar{f}^d}
&\frac{f_{\alpha_1}^{u*}f_{\alpha_1}^d+f_{\alpha_2}^{u*}f_{\alpha_2}^d}
{\bar{f}^u\bar{f}^d}
&-\frac{f_{\alpha_1}^{u*}f_{\alpha_1}^d+f_{\alpha_2}^{u*}f_{\alpha_2}^d}
{\bar{f}^d}\left(\frac{\tilde{m}_{\beta_i}^{u*}\tilde{m}_{\beta_i}^u}
{m^t_0\tilde{m}_{\beta_3}^u}\right)
+\bar{f}^d\frac{\tilde{m}_{\beta_3}^{d*}}{m^b_0} \\[3mm]
\frac{-f_{\alpha_1}^uf_{\alpha_2}^d+f_{\alpha_2}^uf_{\alpha_1}^d}
{\bar{f}^u}
\frac{\tilde{m}_{\beta_i}^{d*}\tilde{m}_{\beta_i}^d}
{m^b_0\tilde{m}_{\beta_3}^d}
&-\frac{f_{\alpha_1}^{u*}f_{\alpha_1}^d+f_{\alpha_2}^{u*}f_{\alpha_2}^d}
{\bar{f}^u}\frac{\tilde{m}_{\beta_i}^{d*}\tilde{m}_{\beta_i}^d}
{m^b_0\tilde{m}_{\beta_3}^d}
+\bar{f}^u\frac{\tilde{m}_{\beta_3}^{u*}}{m^t_0}  & 1 
\end{array} \right)\,,
\end{equation}
At first sight, we see that the Cabbibo angle \cite{cabbibo} $V_{us}$ 
and $V_{cd}$ are not necessarily small.  However, there are ways to 
achieve the smallness, for example, taking 
$f_{\alpha_1}^q=f_{\alpha_2}^q$.  For simplicity, by taking 
$|f_{\alpha_1}^q|\simeq |f_{\alpha_2}^q|\simeq |f_{\alpha_3}^q|$ and 
$\tilde{m}_{\beta_1}^q\simeq\tilde{m}_{\beta_2}^q\simeq\tilde{m}_{\alpha_3}^q$, 
the quark mixing matrix can be consistent with experimental data.  
Furthermore, from Eq. (\ref{3.12}) we explicitly obtain 
\begin{equation}
V_{us} = -V_{cd}^*\,,~~~V_{ub} \sim -10^{-2}V_{us}\,,
~~~V_{td} \sim 10^{-2}V_{us}\,,~~~
\frac{V_{ub}}{V_{td}} \simeq {\mathcal O}(1)\,,
~~~V_{cb} \sim V_{ts} \sim {\mathcal O}(10^{-2})\,.
\end{equation}

The first generation quarks get their masses from the sneutrino VEVs 
\cite{liu2}.  Coming back to Eq. (\ref{3.5}), we see the quantities 
$\lambda'_{u\beta} v_{l_e}$ and $\lambda'_{c\beta} v_{l_{\mu}}$ have 
been implicitly taken to be smaller than the masses of the second family.  
However, they will produce a mass to the down quark of the first family.  
By assuming $\lambda'_{u\beta}$ and $\lambda'_{c\beta}\sim 10^{-2}$, its 
value can be several MeV numerically.  In addition, the mass of the up 
quark of the first family cannot be produced in this way.  This gives us 
an explanation of the fact that $m_d > m_u$, and even may bring us a 
solution to the strong CP problem \cite{liu2}.  We should note that in 
order to keep Eq. (\ref{3.7}) factorisable, that is to make the trilinear 
soft terms to be the mass origin solely for the second generation, we 
have assumed $m_{\tilde{q}^c}^2 \ll m_{\tilde{q}}^2$.  A nonvanishing 
quantity $m_{\tilde{q}^c}^2/m_{\tilde{q}}^2$ contributes a mass to the 
first generation of quarks 
$\sim m_{c(s)}\frac{m_{\tilde{q}^c}^2}{m_{\tilde{q}}^2}$.  For our 
above picture being valid, $m_{\tilde{q}^c}^2/m_{\tilde{q}}^2$ should be 
smaller than $10^{-3}$.  

CP violation originates from the SUSY soft breaking part.  In general, 
there are several possible origins of CP violation within the framework 
of SUSY.  The first one is the complex Yukawa couplings $y_t$ and $y_b$ 
in Eq. (\ref{3.5}) from which, it is seen explicitly that their phases 
can be absorbed by the redefinition of the quark fields.  The second 
possible origin is from the R-parity violating couplings $\lambda'$'s.  
Their CP violation effect is suppressed by the heavy squarks.  As for 
the small down quark mass terms $(\lambda' v_{l_{\alpha}})_{c\beta}$, 
and $(\lambda' v_{l_{\alpha}})_{u\beta}$, not only can most of their 
phases be rotated away, but also are they themselves very small numbers 
($m_d\ll m_s$) in the mass matrix.  Therefore, the R-parity violating 
terms are also not the source of the observed CP violation.  In 
addition, due to the sneutrino VEV can be complex, the sneutrino 
exchange makes CP violating processes.  However, these processes are 
suppressed by the heavy sneutrino.  The third origin lies in the phases 
of the soft breaking terms.  Such an origin is a specific feature of 
SUSY theories.  These phases would in turn enter the quark 
mixing matrix through the radiative mass generation mechanism 
Eq. (\ref{3.7}).  We note that the experimental data of the neutron 
electric dipole moment does not require these phases to be small in 
this model, because its SUSY correction is suppressed by the heavy 
squarks.  From Eq. (\ref{3.13}), we see that the 
Kobayashi-Maskawa (KM) CP violation mechanism \cite{km} can be realized.  
It is the third origin that is the reason CP violation occurs in our 
model.  

\section{The effective theory and the Higgs mass}

The light Higgs is the following combination, 
\begin{equation}
\label{25}
h = a_u h_u + a_d h_d^* + a_e \tilde{l}_e^* 
+ a_{\mu} \tilde{l}_{\mu}^* + a_{\tau} \tilde{l}_{\tau}^*\,, 
\end{equation}
where 
\begin{equation}
\label{26}
a_u = \frac{v_u}{v}\,,~~a_d = \frac{v_d}{v}\,,~~
a_{\alpha} = \frac{v_{l_{\alpha}}}{v}\,, 
\end{equation}
where $v \equiv \sqrt{v_u^2+v_d^2+\sum_{\alpha}v_{l_{\alpha}}^2}$.  
The low energy  effective theory is written as 
\begin{equation}
\label{27}
\begin{array}{lll}
{\mathcal L}_{\rm eff}&=&y_{\tau\tau} l_{\tau} h^{\dag} e^c_{\tau} 
    + y_{\mu\mu} l_{\mu} h^{\dag} e^c_{\mu} 
    + y_{\mu\tau} l_{\mu} h^{\dag} e^c_{\tau} 
    + y_{e\tau} l_e h^{\dag} e^c_{\tau} 
    + y_{e\mu} l_e h^{\dag} e^c_{\mu} \\[3mm]  
& & + y_e^{\alpha\beta} l_{\alpha} h^{\dag} e^c_{\beta}
+\displaystyle\frac{a^2v_{l_{\alpha}}v_{l_{\beta}}}{M_{\tilde{Z}}}
      \nu_{\alpha}^{Tc}\nu_{\beta} \\[3mm]
& & + y_{tt} q_t h t^c + y_{tb} q_t h^{\dag} b^c 
+ y_{cc}^{\alpha\beta} q_{c_{\alpha}} h c_{\beta}^c 
+ y_{cs}^{\alpha\beta} q_{c_{\alpha}} h^{\dag} s_{\beta}^c \\[3mm] 
& & + m^2 h^{\dag}h-\displaystyle\frac{\lambda}{2}(h^{\dag}h)^2+h.c. \,, 
\end{array}
\end{equation}
where the effective Yukawa couplings are 
\begin{equation}
\label{28}
\begin{array}{lll}
y_{\tau\tau} & = & y_{\tau} a_d \,, ~ 
y_{\mu\mu} = \lambda_{\mu} a_e \,, ~
y_{\mu\tau} = \lambda_{\tau} a_e \,, ~
y_{e\tau} = \lambda_{\tau} a_{\mu} \,, ~
y_{e\mu} = \lambda_{\mu} a_{\mu} \,, ~ 
y_e^{\alpha\beta}=\displaystyle\frac{\delta M^l_{\alpha\beta}}{v}\\[3mm]
y_{tt} & = & y_t a_u \,, ~ 
y_{tb} = y_b a_d \,, ~ 
y_{cc}^{\alpha\beta}=\displaystyle\frac{\delta M^u_{\alpha\beta}}{v}\,,~
y_{cs}^{\alpha\beta}=
\displaystyle\frac{\delta M^d_{\alpha\beta}}{v}\,, 
\end{array}
\end{equation}
and $\lambda$ is determined by the gauge couplings, 
\begin{equation}
\lambda = \displaystyle \frac{a^2}{2}
(a_u^*a_u-a_d^*a_d-a_{\alpha}^*a_{\alpha})\,.
\end{equation}

The above quantities are given at a high energy scale which is 
$m_S\simeq 10^{11}$ GeV.  At the EW scale, their values 
can be calculated via the renormalization group method.  It is expected 
that for most of them, the modification is not significant.  We put 
such a systematic analysis for future works.  Nevertheless, the Higgs 
mass should be discussed.  As far as this point is concerned, our model 
is the same as that given in Ref. \cite{axion}.  By taking 
$\tan\beta \sim m_t/m_b$, it was shown \cite{axion} that 
\begin{equation}
m_h \simeq 145 \pm 7 \mbox{ GeV}\,,
\end{equation}
where the uncertainty includes that of both $m_t$ and $\alpha_s$.  

\section{Neutrino oscillation}  

Implications of the neutrino oscillations to this model need a more 
detailed study.  The neutrino masses and lepton mixing deserve a 
separate section to be discussed.  Eq. (\ref{20}) would be a kind of 
democratic mass matrix \cite{democ} for neutrinos 
if $v_{l_e}\sim v_{l_{\mu}}\sim v_{l_{\tau}}$.  The large mixing of the 
neutrinos would seem to be naturally accommodated.  However, this would 
result in a large $\nu_e$-$\nu_{\tau}$ mixing.  In addition, as we have 
mentioned, right-handed neutrinos are needed for the realistic neutrino 
oscillations.  

A SM singlet superfield $N$ is introduced.  In the superpotential, the 
following terms should be added, 
\begin{equation}
\label{a1}
{\mathcal W} \supset \frac{\tilde{\kappa}_1}{\sqrt{3}}\sum_i L_i H_1N 
+ \tilde{M}_1^2 N + \tilde{M}_2NN + \tilde{\kappa}_2H_1H_2N 
+ \tilde{\kappa}_3 N^3 
\end{equation}
with $\tilde{\kappa}_i$'s being coupling constants and $\tilde{M}_i$'s 
masses supposed to be large.  The linear term can be removed away via 
field redefinition.  By defining $\bar{N}=N+n_0$ with $n_0$ being a 
constant field, we write 
\begin{equation}
\label{a2}
\tilde{M}_1^2 N+\tilde{M}_2 N^2+\tilde{\kappa}_3 N^3 = 
\tilde{M} \bar{N}^2 + \tilde{\kappa}_3 {\bar N}^3 + C \,,
\end{equation}
where $\tilde{M}$, $n_0$ and $C$ satisfy 
\begin{equation}
\label{a3}
\begin{array}{lll}
\tilde{M} + 3 \tilde{\kappa}_3 n_0 & = & \tilde{M}_2\,,\\
(2\tilde{M} + 3 \tilde{\kappa}_3 n_0 )n_0 & = & \tilde{M}_1^2 \,, \\
C & = & -\tilde{M} n_0^2 - \tilde{\kappa}_3 n_0^3 \,.  
\end{array}
\end{equation}
In terms of $\bar{N}$, 
\begin{equation}
\label{a4}
{\mathcal W}\supset\frac{\tilde{\kappa}_1}{\sqrt{3}}\sum_iL_iH_1\bar{N}
+\tilde{M}\bar{N}\bar{N}+\tilde{\kappa}_2H_1H_2\bar{N}
+\tilde{\kappa}_3 \bar{N}^3 \,,
\end{equation}
where the constant $C$ is omitted.  Note that the field redefinition 
adds $\tilde{\kappa}_1 n_0/\sqrt{3}$ and $\tilde{\kappa}_2 n_0$ to 
$\tilde{\mu}'$ and $\tilde{\mu}$ in Eq. (\ref{2}), respectively.  These 
are simply regarded as redefinition of $\tilde{\mu}'$ and $\tilde{\mu}$.  
Generally the corresponding soft terms can be written down.  We assume 
that $\bar{N}$ does not develop any non-vanishing VEV.  (There are other 
ways to eliminate the purely linear term of $N$ with a large mass-squared 
coefficient.  For example, $N$ is assumed being charged under a larger 
gauge group than the SM \cite{witten}.  The soft mass of $N$ is assumed 
to be large enough that $N$ has no any non-vanishing VEV.)  Through the 
previous field redefinition, Eq. (\ref{a4}) then becomes 
\begin{equation}
\label{a5}
{\mathcal W} \supset \kappa_{\tau} H_uL_{\tau}\bar{N} 
+ \tilde{M}\bar{N}\bar{N}+\kappa_d H_uH_d\bar{N} 
+\tilde{\kappa}_3 \bar{N}^3 \,,
\end{equation}
where 
\begin{equation}
\label{a6}
\begin{array}{ccc}
\kappa_{\tau} & = & \displaystyle \frac{1}{2\bar{\mu}}
\left[\left(\frac{\mu'}{c_1}-\frac{\mu}{c'_1}\right)\tilde{\kappa}_2
-\left(\frac{\mu'}{c_2}+\frac{\mu}{c'_1}\right)
\frac{\tilde{\kappa_1}}{c_1}\right]\,,\\[5mm]
\kappa_d & = & \displaystyle \frac{1}{2\bar{\mu}}\left[\left(
\frac{\mu'}{c'_1}-\frac{\mu}{c_2}\right)\frac{\tilde{\kappa_1}}{c_1}
+\left(\frac{\mu}{c_1}+\frac{\mu'}{c'_1}\right)\tilde{\kappa}_2\right]\,.
\end{array}
\end{equation}
The last two terms in Eq. (\ref{a5}) do not play important roles to our 
analysis.  The first two terms contribute a mass term to the neutrino 
mass matrix by the seesaw mechanism, 
\begin{equation}
\label{a7}
m_{\nu_{\tau}\nu_{\tau}}\simeq-\frac{(\kappa_{\tau}v_u)^2}{\tilde{M}}\,.
\end{equation}
From Eq. (\ref{20}) or (\ref{27}) and the above mass term, the full 
neutrino mass matrix is 
\begin{equation}
\label{a8}
\begin{array}{lll}
{\mathcal M}^{\nu} & = & \displaystyle - \frac{a^2}{M_{\tilde{Z}}}
\left(
\begin{array}{ccc}
v_{l_e}v_{l_e}      &v_{l_e}v_{l_{\mu}}      &v_{l_e}v_{l_{\tau}}     \\
v_{l_{\mu}}v_{l_e}  &v_{l_{\mu}}v_{l_{\mu}}  &v_{l_{\mu}}v_{l_{\tau}} \\
v_{l_{\tau}}v_{l_e} &v_{l_{\tau}}v_{l_{\mu}} &v_{l_{\tau}}v_{l_{\tau}}+x
\end{array}
\right) 
\end{array}
\end{equation}
with $x$ being $\displaystyle
\frac{M_{\tilde{Z}}}{\tilde{M}}\left(\frac{\kappa_{\tau}v_u}{a}\right)^2$.  
We find that realistic lepton physics can be obtained by taking 
$x \sim v_{l_{\tau}}v_{l_{\tau}}\gg v_{l_e}^2+v_{l_{\mu}}^2 \sim 1$ 
GeV$^2$.  The eigen values are    
\begin{equation}
\label{a9}
\begin{array}{lll}
m_{\nu_3}  & \simeq & \displaystyle \frac{a^2}{M_{\tilde{Z}}} 
v_{l_{\tau}}^2 + \frac{(\kappa_{\tau}v_u)^2}{\tilde{M}} \,, \\[3mm]
m_{\nu_2}  & \simeq & \displaystyle \frac{a^2}{M_{\tilde{Z}}} 
(v_{l_e}^2+v_{l_{\mu}}^2)\frac{x}{x+v_{l_{\tau}}^2} \,, \\[3mm]
m_{\nu_1}  & =      & 0                        \,.
\end{array}
\end{equation}
The solar neutrino problem requires that 
$m_{\nu_2}\simeq (10^{-2}-10^{-3})$ eV which is achieved when 
$M_{\tilde{Z}}\sim 10^{11}$ GeV.  Suppose $v_{l_{\tau}}\sim 10$ GeV, 
the atmospheric neutrino problem requires a certain cancellation between 
the terms $\displaystyle\frac{a^2}{M_{\tilde{Z}}}v_{l_{\tau}}^2$ and 
$\displaystyle\frac{(\kappa_{\tau}v_u)^2}{\tilde{M}}$, in order to make 
$m_{\nu_3}\sim 10 ^{-1}-10^{-2}$.  The mass matrix Eq. (\ref{a8}) is 
diagonalized by $U_{\nu}$, 
\begin{equation}
\label{a10}
U_{\nu} = \left(
\begin{array}{ccc}
\displaystyle \frac{v_{l_{\mu}}}{\sqrt{v_{l_e}^2+v_{l_{\mu}}^2}}      & 
\displaystyle \frac{v_{l_e}}{\sqrt{v_{l_e}^2+v_{l_{\mu}}^2}}          &
\displaystyle \frac{v_{l_e}}{\sqrt{v_{l_{\alpha}}v_{l_{\alpha}}}}     \\
\displaystyle \frac{-v_{l_e}}{\sqrt{v_{l_e}^2+v_{l_{\mu}}^2}}         &
\displaystyle \frac{v_{l_{\mu}}}{\sqrt{v_{l_e}^2+v_{l_{\mu}}^2}}      &
\displaystyle \frac{v_{l_{\mu}}}{\sqrt{v_{l_{\alpha}}v_{l_{\alpha}}}} \\
0 &
\displaystyle
-\frac{\sqrt{v_{l_e}^2+v_{l_{\mu}}^2}}{v_{l_{\tau}}+x/v_{l_{\tau}}} & 
\displaystyle \frac{v_{l_{\tau}}}{\sqrt{v_{l_{\alpha}}v_{l_{\alpha}}}}
\end{array}
\right) \,.
\end{equation}

Going back to the charged lepton masses, the mass matrix is seen from 
Eq. (\ref{12}) or (\ref{27}) - (\ref{28}), 
\begin{equation}
\label{a11}
\begin{array}{lll}
{\mathcal M}^l & = & \left(
\begin{array}{ccc}
0 & \lambda_{\mu} v_{l_{\mu}} & \lambda_{\tau} v_{l_{\mu}} \\
0 & \lambda_{\mu} v_{l_e}     & \lambda_{\tau} v_{l_e}     \\
0 & 0                    & y_{\tau} v_d
\end{array}
\right) \,.
\end{array}
\end{equation}
To obtain agreement with the experimental data, we assume that 
$y_{\tau}v_d\sim\lambda_{\tau}v_{l_{\mu}}\sim\lambda_{\tau}v_{l_e}\sim 1$ 
GeV.  In this case, the mass eigenvalues are 
\begin{equation}
\label{a12}
\begin{array}{lll}
m_{\tau} & \simeq & \sqrt{y_{\tau}^2 v_d^2 + 
|\lambda_{\tau}|^2(v_{l_e}^2+v_{l_{\mu}}^2)} \,,\\ 
m_{\mu}  & \simeq & |\lambda_{\mu}| \sqrt{v_{l_e}^2+v_{l_{\mu}}^2}
\displaystyle \frac{y_{\tau} v_d}{\sqrt{y_{\tau}^2 v_d^2 + 
|\lambda_{\tau}|^2(v_{l_e}^2+v_{l_{\mu}}^2)}} \,,\\
m_e      & =      & 0                        \,.
\end{array}
\end{equation}
The mass-squared matrix ${\mathcal M}^l{\mathcal M}^{l\dagger}$ is 
diagonalized by $U_l$, 
\begin{equation}
\label{a13}
U_l = \left(
\begin{array}{ccc}
\displaystyle \frac{-v_{l_e}}{\sqrt{v_{l_e}^2+v_{l_{\mu}}^2}}    &
\displaystyle \frac{v_{l_{\mu}}}{\sqrt{v_{l_e}^2+v_{l_{\mu}}^2}}
\frac{y_{\tau} v_d}
{\sqrt{y_{\tau}^2 v_d^2+|\lambda_{\tau}|^2(v_{l_e}^2+v_{l_{\mu}}^2)}}& 
\displaystyle \frac{\lambda_{\tau} v_{l_{\mu}}}
{\sqrt{y_{\tau}^2 v_d^2+|\lambda_{\tau}|^2(v_{l_e}^2+v_{l_{\mu}}^2)}}\\
\displaystyle \frac{v_{l_{\mu}}}{\sqrt{v_{l_e}^2+v_{l_{\mu}}^2}} &
\displaystyle \frac{v_{l_e}}{\sqrt{v_{l_e}^2+v_{l_{\mu}}^2}}
\frac{y_{\tau} v_d}
{\sqrt{y_{\tau}^2 v_d^2+|\lambda_{\tau}|^2(v_{l_e}^2+v_{l_{\mu}}^2)}}&
\displaystyle \frac{\lambda_{\tau} v_{l_e}}
{\sqrt{y_{\tau}^2 v_d^2+|\lambda_{\tau}|^2(v_{l_e}^2+v_{l_{\mu}}^2)}}\\
0 &
\displaystyle \frac{-\lambda_{\tau}^*\sqrt{v_{l_e}^2+v_{l_{\mu}}^2}}
{\sqrt{y_{\tau}^2 v_d^2+|\lambda_{\tau}|^2(v_{l_e}^2+v_{l_{\mu}}^2)}}&
\displaystyle \frac{y_{\tau}v_d}
{\sqrt{y_{\tau}^2 v_d^2+|\lambda_{\tau}|^2(v_{l_e}^2+v_{l_{\mu}}^2)}}
\end{array}
\right) \,.
\end{equation}

The lepton mixing matrix is $V \equiv U_l^{\dagger} U_{\nu}$.  The 
$\nu_e$-$\nu_{\mu}$ mixing is 
\begin{equation}
\label{a14}
|V_{e2}| = \frac{v_{l_{\mu}}^2-v_{l_e}^2}{v_{l_e}^2+v_{l_{\mu}}^2} \,.
\end{equation}
It is ${\mathcal O}(1)$ by taking $v_{l_e} \sim v_{l_{\mu}}$.  The 
$\nu_{\mu}$-$\nu_{\tau}$ mixing is 
\begin{equation}
\label{a15}
|V_{\mu3}| \simeq \frac{|\lambda_{\tau}|\sqrt{v_{l_e}^2+v_{l_{\mu}}^2}}
{\sqrt{y_{\tau}^2 v_d^2+|\lambda_{\tau}|^2(v_{l_e}^2+v_{l_{\mu}}^2)}}\,.
\end{equation}
A maximal mixing is approached if $y_{\tau}v_d$ is getting equal to  
$|\lambda_{\tau}|\sqrt{v_{l_e}^2+v_{l_{\mu}}^2}$. (That is the reason we 
have assumed a large $\lambda_{\tau}\sim {\mathcal O}(1)$.)  The 
current data show that $|V_{\mu3}|$ can be as small as $0.6$ at the 
$99\%$ C.L. \cite{sv}.  Finally the $\nu_e$-$\nu_{\tau}$ mixing is 
\begin{equation}
\label{a16}
|V_{e3}| \simeq \frac{v_{l_{\mu}}^2-v_{l_e}^2}
{\sqrt{v_{l_e}^2+v_{l_{\mu}}^2}v_{\tau}} \,.
\end{equation}
It is small $\sim 0.1$ if 
$\sqrt{v_{l_e}^2+v_{l_{\mu}}^2}/v_{\tau} \sim 0.1$.  A generic 
expectation of $|V_{e3}|$ is around $0.1$ which is near to its 
experimental limit $|V_{e3}|< 0.17$ \cite{sv}.  

\section{Discussions}  

Some important aspects of our model should be discussed.  In this 
framework, we do not need to introduce any extra symmetry, like the 
R-parity, or the baryon number, to forbid the proton decay.  The 
interactions of baryon number violation should be included in the 
superpotential in principle, 
\begin{equation}
\label{4.1}
{\mathcal W} \supset \lambda''_{ijk} U^c_i D^c_j D^c_k \,,
\end{equation}
with $\lambda''_{ijk}$ being coupling constants.  Here the analysis is 
essentially the same as that of R-parity violation in split SUSY 
\cite{r-parity2}.  Because the sparticles are very heavy, they suppress 
baryon number and some lepton number violating processes to be 
unobservable, despite that the coupling constants of the baryon and 
lepton number violating interactions are large.  For example, in the 
$\tau\to\mu^+\mu^-e$ decay which occurs at tree level, the branching 
ratio is about $\sim 10^{-23}$ if $m_S\sim 10^{11}$ GeV.  The proton 
decay measurements constraint \cite{proton-decay} 
$\lambda'\lambda''\leq 10^{-27}\displaystyle\frac{m_S^2}{(100 \mbox{ GeV})^2}$. 
When $m_S\sim 10^{11}$ GeV and $\lambda' \sim 10^{-2}$, $\lambda''$ is 
required to be smaller than $10^{-7}$.  

As we have mentioned in Sect. III, $U_i^c$'s can compose a nontrivial 
representation of $Z_{3L}$ symmetry.  In this case, the baryon number 
violating terms in Eq. (\ref{4.1}) is written as 
\begin{equation}
\label{4.2}
{\mathcal W} \supset \sqrt{3} \lambda''_{tjk} U^c_t D^c_j D^c_k \,,
\end{equation}
where $U^c_t=\displaystyle \frac{1}{\sqrt{3}} (U_1^c+U^c_2+U^c_3)$.  
This interaction does not lead to proton decays in the massless 
up-quark case, and $\lambda''$ can be ${\mathcal O}(1)$ numerically.  
Consider the case of a nonvanishing up quark mass, the proton decay rate 
is suppressed by the up-quark and top-quark mass ratio which is 
$(m_u/m_t)^2\sim 10^{-10}$, and $\lambda''$ can be $10^{-2}$.  

From the EW symmetry breaking point of view, this model is nothing but a 
fine-tuned minimal SUSY SM with a high SUSY breaking scale.  
In spite of losing naturalness of the EW energy scale, the radiative
breaking mechanism for the EW gauge symmetry \cite{susy2,susy3} may 
still remain in this model, because the situation is similar to that in 
the minimal SUSY SM.  This SUSY model does not suffer from the so-called
$\mu$-problem \cite{mu}.  Throughout the analysis, 
$\tilde{\mu}^{(\prime)}$ in Eq. (\ref{2}) are not necessarily required 
to have the same order as that of the soft masses.  They can be several 
orders smaller than the soft masses.  Furthermore even the soft masses 
are not required to be at the same order.  Because these masses are much 
larger than the EW scale, their differences do not cause any 
inconsistency phenomenologically.  It seems that the unification of the 
gauge coupling constants is lost.  There are arguments \cite{calmet} 
that the unification can be still true in the case of high scale SUSY 
breaking.  Nevertheless, the general trend of approximate unification is 
still there.  This model does not suffer from the second fine-tuning of 
the split SUSY \cite{manuel}, because SUSY has no huge split in this 
model.

It would be difficult for experiments to verify the model directly, 
except for the $145$ GeV $m_h$ and the ${\mathcal O}(0.1)$ $\theta_{13}$.  
At low energies, the model is basically the same as the SM.   
One essential feature of this model is that the unnaturally light 
Higgs has a component of a slepton.  Related to this point, the model 
allows for relatively long-lived Higgsinos.  We may consider a case 
where their masses are lower than $m_S$.  If they are loop induced, 
the Higgsino masses are thousand times smaller than $m_S$.  A Higgsino 
decays to a Higgs and a virtual gaugino which further goes into a 
lepton and a virtual slepton, the slepton decays to a lepton pair via 
R-parity violating interaction (Fig. 3).  Because this four body decay 
is suppressed by the R-parity violating coupling and double suppressed 
by $m_S$, a $10^8$ GeV heavy Higgsino has a lifetime of 
$10^{-5}$ sec.  

\begin{figure}
\includegraphics{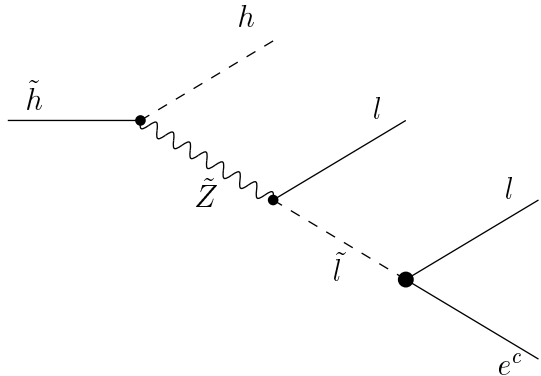}
\caption{\label{fig3}
A Higgsino decays to a Higgs and a virtual gaugino which further goes 
into a lepton and a virtual slepton, the slepton decays to a lepton 
pair via R-parity violating interaction.}
\end{figure}

The cosmological and astrophysical implications should be studied in 
future works.  CP violation at high energies $\geq m_S$ has various 
origins, like the phases in soft SUSY breaking terms, in R-parity 
violating couplings, and in the sneutrino VEV.  They might be the root 
of the matter-antimatter asymmetry of the Universe.  However at the EW 
scale, many of these sources are suppressed.  What left is reduced to 
the KM mechanism.  On the other hand, we have noted that there is no 
natural dark matter candidate in this model.  Existence of the dark 
matter implies that the dark matter is really dark, namely it only has 
gravity interaction with the ordinary matter.  

It should be studied in future works how the soft 
SUSY breaking terms also break the $Z_{3L}$ family symmetry.  
Currently we understand this as follows.  The gravity breaks any global 
symmetry like our family symmetry.  The superpotential is for particle 
physics, which decouples from the gravity, and therefore keeps the 
family symmetry.  Whereas the soft SUSY breaking terms are due to 
gravitational interaction, they may violate the family symmetry 
explicitly.  
This model would prefer a large $\nu_e$-$\nu_{\tau}$ mixing, if we had 
not looked at the experimental data for the leptons.  A large 
$\lambda_{\tau} \sim 1$ is assumed in order to fit the data.  In 
addition, a special structure of the soft breaking terms of the squark 
was assumed to produce the second and first generation quark mass 
hierarchy.  The problem that how natural these assumption are also needs 
further studies.  

There are various particle physics models which do not care the 
naturalness problem of the SM.  This model seems to be one of them.  
However, this model is unique in the sense that it makes use of a 
mechanism which used to be directly for naturalness of the SM.  In 
other words, while many other models can be naturalized after their 
SUSY extension, this model is intrinsically unnatural.  Unnaturalness 
is understood by the anthropic principle.  One meaningful question is 
then that why are not the SM Yukawa couplings anthropically determined?  
We answer this question from the following aspects.  One is that 
although the electron mass might be understood from the anthropic 
point of view, it is hard to say all the Yukawa couplings including 
the neutrino masses being anthropically determined.  Theoretically we 
have given up naturalness in the t'Hooft sense.  Rather, Dirac 
naturalness has been emphasized in considering the fermion mass 
hierarchies.  Whether such an effort can be justified should be studied 
in the more fundamental theory.  

Finally we make a remark on the possibility of lowering the SUSY 
breaking scale from $10^{11}$ GeV to $1$ TeV.  This work essentially 
has made use of SUSY to understand fermion masses.  
The SUSY breaking scale is fixed by the neutrino masses implied by 
the neutrino oscillations.  It would be much more interesting 
if SUSY also plays the role of stabilizing the EW scale \cite{susy1}.  
We note that it is still possible that the $\tau$-neutrino mass is about 
$10$ MeV \cite{pdg,liu3,ls} (in this case the atmospheric neutrino 
problem is due to $\nu_{\mu}-\nu_{\rm sterile}$ oscillation which is 
also yet ruled out experimentally), then the SUSY breaking scale will be 
about a few TeV.  In that case the EW energy scale will be marginally 
natural.  

\section{Summary}

If SUSY is not for stabilizing the EW energy scale, what is it used for 
in particle physics?  In this paper, motivated by our previous works 
\cite{liu1,liu2,liu4}, we have proposed that SUSY is for flavor problems.  
A family symmetry $Z_{3L}$, which is the cyclic symmetry among the three 
generation $SU(2)_L$ doublets, is introduced.  No additional global 
symmetry, like the R-parity is imposed.  SUSY breaks at a high scale 
$\sim 10^{11}$ GeV.  The EW energy scale $\sim 100$ GeV is unnaturally 
small from the point of view of the field theory.  Under the family 
symmetry, only the third generation fermions get to be massive after EW 
symmetry breaking.  This family symmetry is broken by soft SUSY breaking 
terms.  These terms contribute masses via loops to the second generation 
quarks and the electron.  Furthermore they induce sneutrino VEVs which 
result in the masses of the muon and the down quark.  The neutrino large 
mixing can be obtained.  The KM mechanism of CP violation is 
realized at low energies.  A hierarchical pattern of the lepton and quark 
masses are obtained.  The Higgs mass of this model is about $145$ GeV.  
This point can be tested in the future experiments at Tevatron and LHC.   
It is expected that $\nu_e$-$\nu_{\tau}$ mixing is near to its 
experimental limit.  

\begin{acknowledgments}
We would like to thank Yue-Liang Wu, Jin Min Yang, Zhi-Zhong Xing and 
Pyungwon Ko for helpful discussions.  The author acknowledges support 
from the National Natural Science Foundation of China.
\end{acknowledgments}


\newpage 

\begin{thebibliography}{99}

\bibitem{susy}
J. Wess and B. Zumino, Nucl. Phys. B 70 (1974) 39;\\
Y. Gol'fand and E. Likhtman, JETP Lett. 13 (1971) 323;\\
D.V. Volkov and V. Akulov, Phys. Lett. B 46 (1973) 109.

\bibitem{susy1}
E. Witten, Nucl. Phys. B 188 (1981) 513;\\
S. Dimopoulos and H. Georgi, Nucl. Phys. B 193, 150 (1981);\\
N. Sakai, Z. Phys. C 11 (1981) 153.

\bibitem{susy2}
L. Ibanez and G.G. Ross, Phys. Lett. B 110 (1982) 215;\\
L. Alvarez-Gaume, M. Claudson and M.B. Wise, Nucl. Phys. B 207 (1982) 96.  

\bibitem{gut}
J.C. Pati and A. Salam, Phys. Rev. D 10 (1974)275;\\
H. Georgi and S.L. Glashow, Phys. Rev. Lett. 32 (1974) 438.

\bibitem{gut1}
P. Langacker, M.-X. Luo, Phys. Rev. D 44 (1991) 817;\\
C. Giunti, C.W. Kim, U.W. Lee, Mod. Phys. Lett. A 6 (1991) 1745;\\ 
U. Amaldi, W. de Boer, H. Furstenau, Phys. Lett. B 260 (1991) 447;\\ 
J. Ellis, S. Kelley, D. Nanopolous, Phys. Lett. B 260 (1991) 131.

\bibitem{cc}
S. Weinberg, Phys. Rev. Lett. 59 (1987) 2607.  

\bibitem{string}
R. Bousso and J. Polchinski, JHEP 0006 (2000) 006; 
S. Karchru, R. Kallosh, A.D. Linde and S.P. Trivedi, 
Phys. Rev. D 68 (2003) 046005; 
S.B. Giddings, S. Karchru and J. Polchinski, 
Phys. Rev. D 66 (2002) 106006; 
A. Maloney, E. Silverstein and A. Strominger, hep-th/0205316.  

\bibitem{sm}
V. Agrawal, S.M. Barr, J.F. Donoghue and D. Seckel,
Phys. Rev. D 57 (1998) 5480. 

\bibitem{split}
N. Arkani-Hamed and S. Dimopoulos, 
JHEP 0506 (2005) 073 [hep-th/0405159].

\bibitem{split1}
G. F. Giudice and A. Romanino, 
Nucl. Phys. B 699 (2004) 65 [hep-ph/0406088].  

\bibitem{split2}
N.~Arkani-Hamed, S.~Dimopoulos, G.~F.~Giudice and A.~Romanino,
Nucl. Phys. B 709 (2005) 3 [hep-ph/0409232]; 
A.~Arvanitaki, C.~Davis, P.~W.~Graham and J.~G.~Wacker,
Phys. Rev. D 70 (2004) 117703 [hep-ph/0406034]; 
A.~Pierce,
Phys. Rev. D 70 (2004) 075006 [hep-ph/0406144]; 
S.~h.~Zhu,
Phys. Lett. B 604 (2004) 207 [hep-ph/0407072];
B.~Mukhopadhyaya and S.~SenGupta,
Phys. Rev. D 71 (2005) 035004 [hep-th/0407225]; 
W. Kilian, T. Plehn, P. Richardson, E. Schmidt, 
Eur.Phys.J. C39 (2005) 229 [hep-ph/0408088]; 
R.~Mahbubani,
hep-ph/0408096; 
M.~Binger,
Phys. Rev. D 73 (2006) 095001 [hep-ph/0408240]; 
J.~L.~Hewett, B.~Lillie, M.~Masip and T.~G.~Rizzo,
JHEP 0409 (2004) 070 [hep-ph/0408248]; 
L.~Anchordoqui, H.~Goldberg and C.~Nunez,
Phys.Rev. D71 (2005) 065014 [hep-ph/0408284]; 
D.A. Demir, 
hep-ph/0410056; 
R. Allahverdi, A. Jokinen, A. Mazumdar, 
Phys. Rev. D 71 (2005) 043505 [hep-ph/0410169]; 
B.~Bajc and G.~Senjanovic,
Phys. Lett. B 610 (2005) 80 [hep-ph/0411193]; 
B.~Kors and P.~Nath,
Nucl. Phys. B 711 (2005) 112 [hep-th/0411201];
M.A. Diaz and P.F. P\'erez, 
J. Phys. G 31 (2005) 563-569 [hep-ph/0412066]; 
E.~J.~Chun and S.~C.~Park,
JHEP 0501 (2005) 009 [hep-ph/0410242]; 
K.~Cheung and W.~Y.~Keung,
Phys. Rev. D 71 (2005) 015015 [hep-ph/0408335]; 
J.D. Wells, 
Phys. Rev. D71 (2005) 015013 [hep-ph/0411041]; 
A.~Masiero, S.~Profumo and P.~Ullio,
Nucl. Phys. B 712 (2005) 86[ hep-ph/0412058]; 
A. Arvanitaki, P. W. Graham, 
Phys. Rev. D 72 (2005) 055010 [hep-ph/0411376];
L. Senatore, 
Phys. Rev. D 71 (2005) 103510 [hep-ph/0412103]; 
P. C Schuster, 
hep-ph/0412263; 
A. Datta, X.-m. Zhang, 
Int. J. Mod. Phys. A 21 (2006) 2431 [hep-ph/0412255]; 
J.-j. Cao, J. M. Yang, 
Phys. Rev. D (2005) 111701 [hep-ph/0412315]; 
C.-H. Chen and C.-Q. Geng,
Phys. Rev. D 72 (2005) 037701 [hep-ph/0501001].  

\bibitem{calmet}
X. Calmet, 
Eur. Phys. J. C 41 (2005) 245 [hep-ph/0406314];\\
V. Barger, J. Jiang, P. Langacker and T.-j. Li,
Phys. Lett. B 624 (2005) 233 [hep-ph/0503226].  

\bibitem{axion}
V. Barger, C.-W. Chiang, J. Jiang, T.-j. Li, 
Nucl. Phys. B 705 (2005) 71 [hep-ph/0410252].  

\bibitem{manuel}
M. Drees, 
hep-ph/0501106.  

\bibitem{r-parity2}
S.K. Gupta, P. Konar, B. Mukhopadhyaya, 
Phys. Lett. B 606 (2005) 384 [hep-ph/0408296];\\
P.F. P\'erez, 
J. Phys. G 31 (2005) 1025-1030 [hep-ph/0412347].  

\bibitem{liu4}
C. Liu, 
Phys. Lett. B 609 (2005) 111. [hep-ph/0501129]

\bibitem{he}
For a recent study, see D.A. Dicus and H.-J. He, 
Phys. Rev. D 71 (2005) 093009 [hep-ph/0409131].  

\bibitem{family-symmetry}
H. Fritzsch, Phys. Lett. B 70 (1977) 436; 
S. Adler,
Phys. Rev. D 59 (1999) 015012; Erratum-ibid. D 59 (1999) 099902.
For recent studies, see e.g. 
T. Kitabayashi and M. Yasu\'e, Phys. Rev. D 67 (2003) 015006; 
P. F. Harrison, W. G. Scott, Phys. Lett. B 557 (2003) 76.

\bibitem{liu1}
D. Du and C. Liu, Mod. Phys. Lett. A 8 (1993) 2271; 
A 10 (1995) 1837;\\ 
For a review, see C. Liu, in Beijing 1999, 
Frontier of Theoretical Physics, p. 131 [hep-ph/0005061].

\bibitem{liu2}
C. Liu, Int. J. Mod. Phys. A 11 (1996) 4307.  

\bibitem{r-parity}
C. Aulakh and R. Mohapatra, Phys. Lett. B 119 (1982) 136; \\
S. Weinberg, Phys. Rev. D 26 (1982) 287;\\
For reviews, see G. Bhattacharyya, hep-ph/9709395;\\ 
O.C.W. Kong, Int. J. Mod. Phys. A 19 (2004) 1863;\\ 
M. Chemtob, Prog. Part. Nucl. Phys. 54 (2005) 71.  

\bibitem{r-parity1}
F. Zwirner, Phys. Lett. B 132 (1983) 103;
L. Hall and M. Suzuki, Nucl. Phys. B 231 (1984) 419;
I. H. Lee, Phys. Lett. B 138 (1984) 121;
G. Ross and J. Valle, Phys. Lett. B151 (1985) 375;
J. Ellis et al., Phys. Lett. B 150 (1985) 142;
S. Dawson, Nucl. Phys. B 261 (1985) 297;
R. Barbieri and A. Masiero, Nucl. Phys. B 267 (1986) 679;
S. Dimopoulos and L. Hall, Phys. Lett. B 207 (1988) 210 (1988); 
V.D. Barger, G.F. Giudice and T. Han, 
Phys. Rev. D 40 (1989) 2987;   
C. Liu, Mod. Phys. Lett. A 12 (1997) 329-336; 
G. Bhattacharyya, H.V. Klapdor-Kleingrothaus and H. P\"as,
Phys. Lett. B 463 (1999) 77;
C.-H. Chang and T.-F. Feng,
Eur. Phys. J. C 12 (2000) 137; 
M. Bisset, O.C.W. Kong, C. Macesana and L.H. Orr,
Phys. Rev. D 62 (2000) 035001;
A. Abada and G. Bhattacharyya, 
Phys. Rev. D 63 (2001) 017701;
A.S. Joshipura, R.D. Vaidya and S.K. Vempati, 
Nucl. Phys. B 639 (2002) 290;
Phys. Rev. D 65 (2002) 053018;
A. Abada, S. Davidson and M. Losada, 
Phys. Rev. D 65 (2002) 075010;
A. Abada, G. Bhattacharyya and M. Losada,
Phys. Rev. D 66 (2002) 071701 (R);
Y. Uehara, 
Phys. Lett. B 537 (2002) 256;
M. G\'{o}\'{z}d\'{z}, W.A. Kami\'{n}ski, F. \v{S}imkovic 
and A. Faessler, Phys. Rev. D 74 (2006) 055007.

\bibitem{liu3}
C. Liu and H. S. Song, Nucl. Phys. B 545 (1999) 183.  

\bibitem{seesaw}

P. Minkowski, Phys. Lett. B 67 (1977) 421; 
T. Yanagida, in {\it Proc. of the Workshop on Unified Theory and Baryon
Number of the Universe} (KEK, Tsukuba, 1979), p95; 
M. Gell-Mann, P. Ramond and R. Slansky, in Sanibel talk, CALT-68-709
(Feb. 1979), and in {\it Supergravity} (North Holland, Amsterdam, 1979), 
p315;
S.L. Glashow, in {\it Quarks and Leptons} 
(Plenum, New York, 1980), p. 707;
R. N. Mohapatra and G. Senjanovic, Phys. Rev. Lett. 44 (1980) 912.  

\bibitem{ah}
N. Arkani-Hamed and L. Hall, Phys. Rev. D 54 (1996) 2242-2260; 
Phys. Rev. D 54 (1996) 7032-7050.  

\bibitem{banks}
T. Banks, Nucl. Phys. B 303 (1988) 171.  

\bibitem{cabbibo}
N. Cabibbo, 
Phys. Rev. Lett. 10 (1963) 531;\\
M. Gell-Mann and M. Levy, Nuovo Cimento 16 (1960) 705.

\bibitem{km}
M. Kobayashi and T. Maskawa, 
Prog. Theor. Phys. 49 (1973) 652.

\bibitem{democ}
H. Harari, H. Haut, and J. Weyers, Phys. Lett. B78 (1978) 459;
Y. Chikashige, G. Gelmini, R.P. Peccei, and M. Roncadelli,
Phys. Lett. B94 (1980) 499;
H. Fritzsch, in Proc. of Europhys. Conf. on Flavor Mixing in Weak
Interactions, Erice (1984);
C. Jarlskog, in Proc. of Int. Symp. on Production and Decay of Heavy
Flavors, Heidelberg (1986);
Y. Nambu, in Proc. XI Warsaw Symp. on High Energy Physics, Kazimierz
(1988);
P. Kaus and S. Meshkov, Mod. Phys. Lett. A3 (1988) 1251;
Y. Koide, Z. Phys. C 45 (1989) 39;
M. Tanimoto, Phys. Rev. D 41 (1990) 1586;
G.C. Branco, J.I. Silva-Marcos, and M.N. Rebelo, Phys. Lett. B237
(1990) 446;
H. Fritzsch and J. Plankl, Phys. Lett. B 237 (1990) 451;
D. Du and C. Liu, Mod. Phys. Lett. A 8 (1993) 2271;
H. Fritzsch and Z.Z. Xing, Phys. Lett. B353 (1995) 114;
K. Kang and S.K. Kang, Phys. Rev. D 56 (1997) 1511.

\bibitem{witten}
E. Witten,
Phys. Lett. B 105 (1981) 267.

\bibitem{sv}
A. Strumia and F. Vissani, Nucl. Phys. B 726 (2005) 294 
[hep-ph/0503246].  

\bibitem{proton-decay}
I. Hinchliffe and T. Kaeding, Phys. Rev D 47 (1993) 273;\\
A.Y. Smirnov and F. Vissani, Phys. Lett. B 380 (1996)317.

\bibitem{susy3}
D.M. Pierce, J.A. Bagger, K.T. Matchev and R.-J. Zhang, 
Nucl. Phys. B 491 (1997) 3.  

\bibitem{mu}
J.E. Kim and H.P. Nilles, 
Phys. Lett. B 138 (1984) 150.  

\bibitem{ls}
C. Liu and J.-H. Song, 
Nucl.Phys. B598 (2001) 3.  

\bibitem{pdg}
Particle data group (S. Eidelman et al.), Phys. Lett. B 592 (2004) 1.  

\end{thebibliography}
\end{document}